\documentclass[preprint,12pt]{aastex}
\usepackage{epstopdf}

\newcommand{\vtan}{$V_{tan}$}
\newcommand{\kms}{km~s$^{-1}$}
\newcommand{\ms}{m~s$^{-1}$}
\newcommand{\msun}{M$_{\sun}$}
\newcommand{\ldl}{$\lambda/{\Delta}{\lambda}$}
\newcommand{\lbol}{$\log_{10}{L_{bol}/L_{\sun}}$}

\newcommand{\lii}{\ion{Li}{1}}
\newcommand{\ki}{\ion{K}{1}}
\newcommand{\nai}{\ion{Na}{1}}

\newcommand{\meth}{CH$_4$}

\newcommand{\wat}{H$_2$O}
\newcommand{\name}{2MASS~J20261584$-$2943124}
\newcommand{\namesh}{2MASS~J2026$-$2943}
\newcommand{\chisq}{$\chi^2$}

\slugcomment{Submitted to AJ; draft \today}

\shorttitle{The Unresolved Binary 2MASS~J2026$-$2943}
\shortauthors{Gelino \& Burgasser}

\begin{document}

\title{2MASS~J20261584$-$2943124:\\ An Unresolved L0.5 + T6 Spectral Binary\altaffilmark{1}}

\author{Christopher R.\ Gelino}
\affil{Infrared Processing and Analysis Center, MC 100-22, California Institute of Technology, Pasadena, CA 91125, USA}
\and
\author{Adam J.\ Burgasser\altaffilmark{2,3}}
\affil{Center for Astrophysics and Space Science, University of California San Diego, La Jolla, CA 92093, USA}

\altaffiltext{1}{Some of the data presented herein were obtained at the
  W.M. Keck Observatory, which is operated as a scientific partnership
  among the California Institute of Technology, the University of
  California, and the National Aeronautics and Space Administration. The
  Observatory was made possible by the generous financial support of the
  W.M. Keck Foundation.}
\altaffiltext{2}{Also Massachusetts Institute of Technology, Kavli Institute for Astrophysics and Space Research, Building 37, Room 664B, 77 Massachusetts Avenue, Cambridge, MA 02139, USA.}
\altaffiltext{3}{Visiting Astronomer at the Infrared Telescope Facility, which is operated by
the University of Hawai`i under Cooperative Agreement NCC 5-538 with the National Aeronautics
and Space Administration, Office of Space Science, Planetary Astronomy Program.}

\begin{abstract}
We identify the L dwarf 2MASS~J20261584$-$2943124 
as an unresolved spectral binary, 
based on low-resolution, near-infrared spectroscopy
from IRTF/SpeX.  
The data reveal a peculiar absorption feature at 1.6~$\micron$, previously noted in the spectra of 
other very low-mass spectral binaries, which likely arises from overlapping FeH and {\meth}
absorption bands in the blended light of an L dwarf/T dwarf pair. 
Spectral template matching analysis indicates component types of 
L0.5 and T6, with relative brightness $\Delta{H}$ = 4.2$\pm$0.6.  
Laser guide star adaptive optics imaging observations with Keck/NIRC2
fail to resolve the source, indicating a
maximum separation at the observing epoch of 0$\farcs$25, or a projected separation of 9 AU  
assuming a distance of 36$\pm$5~pc.
With an age that is likely to be relatively older ($\gtrsim$5~Gyr) based on the system's 
large {\vtan} and mass ratio arguments, the relative motion of the 
potentially ``massive'' (0.06-0.08~{\msun}) components of 2MASS~J2026$-$2943 may be detectable through 
radial velocity variations,
like its earlier-type counterpart 2MASS~J03202839$-$0446358 (M8+T5), providing dynamical mass measurements that span the hydrogen burning limit.
\end{abstract}

\keywords{
stars: binaries: general ---
stars: fundamental parameters ---
stars: individual ({\name}) ---
stars: low mass, brown dwarfs
}

\section{Introduction}

Multiple systems are of fundamental importance in
stellar astrophysics and are particularly crucial for studies of very
low mass stars (VLM; M $<$ 0.1 M$_{\odot}$) and brown dwarfs, the M, L, and T
dwarfs (see \citealt{2005ARA&A..43..195K} and references therein). 
The frequency and characteristics of VLM multiples
provide empirical constraints for currently debated formation
mechanisms (e.g., \citealt{2007prpl.conf..427B, 2007prpl.conf..443L}), and are
the dominant outlet for direct mass and radius measurements (e.g., \citealt{2001ApJ...560..390L, 2004ApJ...615..958Z, 2006Natur.440..311S, 2008ApJ...689..436L, 2009ApJ...692..729D}). Resolved
VLM multiples provide useful laboratories for studying
atmospheric properties independent of age and composition effects
(e.g., \citealt{2005ApJ...634..616L, 2007ApJ...657.1064M, 2010ApJ...710.1142B}), and yield unique insights into
evolutionary processes such as the poorly-understood L dwarf/T dwarf transition 
%(e.g., \citealt{2002ApJ...571L.151B, 2004AJ....127.3553K, 2006ApJ...647.1393L, 2006ApJ...640.1063B, 2007ApJ...659..655B, 2008ApJ...678.1372C, 2008ApJ...685.1183L, 2008ApJ...689.1327S}).
(e.g., \citealt{2002AJ....124.1170D,2006ApJ...647.1393L, 2007ApJ...659..655B, 2008ApJ...685.1183L}).

Current VLM multiplicity studies are dominated by resolved imaging
programs, which have uncovered $\sim$100 systems to date\footnote{For an updated list, see \url{http://www.vlmbinaries.org}.}. These
systems have constrained the frequency
($\sim$15-35\%), and separation and mass ratio distributions of VLM multiples. 
Yet resolved imaging studies are insensitive to very
closely-separated ($\rho <$ 1 AU) and/or distant multiples, as well as
multiples observed in unfortunate geometries (e.g., \citealt{1999Sci...283.1718M,2005ApJ...634..616L, 2006PASP..118..611G}). 
The closely-separated systems are particularly valuable, as they are the ones most likely to exhibit measurable orbital motion in a reasonable time frame, and also have the best chance of eclipsing.  Furthermore, there is growing evidence that a significant fraction - perhaps half - of VLM multiples are hidden in unresolved, tight systems, based on searches for radial velocity (RV) variables (e.g., \citealt{2005MNRAS.362L..45M,2006AJ....132..663B,2008A&A...492..545J}) and overluminous sources in cluster color-magnitude diagrams (e.g., \citealt{2003MNRAS.342.1241P,2007MNRAS.380..712L}).  Yet actual yields of very tight binaries have been small due to the inherent inefficiencies associated with these search programs: RV monitoring programs require large allocations of large telescope time, while searches for overluminous sources need clusters and long-term astrometric programs to confirm object membership.  Other signatures of unresolved multiplicity, such as eclipsing (e.g., \citealt{2006Natur.440..311S}), astrometric wobble (e.g., \citealt{2008ApJ...686..548D}) and microlensing (e.g., \citealt{2009arXiv0911.2706B}) have similarly yielded few detections.

%The closely-separated systems are of particular importance,
%potentially contributing substantially to the total VLM binary
%population ($\sim$30-50\% of all binaries; \citealt{2005MNRAS.362L..45M}),
%and providing the opportunity for orbital mass measurements (e.g.,
%\citealt{2008ApJ...678L.125B}) and radius measurements if the system eclipses
%(e.g., \citealt{2006Natur.440..311S}). While spectroscopic
%monitoring have uncovered a handful of these tight pairs
%(e.g., \citealt{1999AJ....118.2460B, 2007ApJ...666L.113J, 2008ApJ...678L.125B}), such programs are generally inefficient, requiring large allocations of
%large telescope time, and they cannot access statistically significant
%samples of intrinsically faint late-type dwarfs.

We have recently developed a technique to identify and characterize 
a subset of VLM ``spectral'' binaries, or blended-light pairs, using low-resolution, near-infrared spectroscopy \citep{2007AJ....134.1330B, 2008ApJ...681..579B, 2010ApJ...710.1142B}.
This technique exploits the distinct and
complex structure of late-type M, L, and T dwarf near-infrared spectra as shaped by absorption bands of {\wat}, {\meth}, CO, and FeH, as well as condensate cloud opacity
(e.g., \citealt{2001AJ....121.1710R, 2003ApJ...596..561M, 2005ApJ...623.1115C}).  Unresolved pairs with different component spectral types exhibit specific spectral peculiarities which arise from blended features.  These peculiarities serve to both identify a source as a spectral binary and enable the decomposition of the spectrum into its component types.  This technique is not limited by the angular separation of a binary, so closely-separated systems---for which dynamic mass measurements and possibly radius measurements are feasible---can potentially be identified.

An illustrative case is 2MASS~J03202839$-$0446358 (hereafter 2MASS~J0320$-$0446; \citealt{2003IAUS..211..197W}), whose near-infrared spectrum exhibits an unusual absorption feature at 1.6~$\mu$m not seen in normal M or L dwarf spectra.  \citet{2008ApJ...681..579B} were able to reproduce the spectrum of this source as the combination of an M8 primary plus T5 secondary, with the 1.6~$\mu$m feature arising from overlapping FeH and {\meth} absorption from the two components, respectively.  2MASS~J0320$-$0446 was independently identified as a spectroscopic binary with an 8-month orbit by \citet{2008ApJ...678L.125B} from high-resolution spectroscopic monitoring.  The inferred separation and system mass function from the Blake et al.\ study, and the component spectral types from the Burgasser et al.\ study, have enabled a firm constraint on the minimum age and component masses of 2MASS~J0320$-$0446 ($>$2~Gyr; M$_1 >$ 0.080~{\msun}, M$_2 >$ 0.053~{\msun}; \citealt{2009AJ....137.4621B}).  
To date, 20 late-type M/L plus T dwarf spectral binaries have been identified in this manner \citep{2004ApJ...604L..61C, 2007AJ....134.1330B, 2008ApJ...681..579B, 2010ApJ...710.1142B, 2008ApJ...685.1183L, 2008arXiv0811.0556S}, three of which have thus far been resolved (\citealt{2006ApJS..166..585B}; C.\ Gelino et al. 2010, in preparation).

This article reports the discovery of a new L dwarf plus T dwarf spectral binary
identified from low-resolution, near-infrared spectroscopy, 
{\name} (hereafter {\namesh}; \citealt{2007AJ....133..439C}).
The spectrum of this source shows the distinct signature of blended FeH and {\meth} absorption, but is  
unresolved in high angular resolution imaging. 
In Section~2 we describe our spectroscopic observations of {\namesh},
obtained with the NASA Infrared Telescope Facility (IRTF) 
SpeX spectrograph \citep{2003PASP..115..362R},
and present our template fitting analysis that demonstrates the binary nature
of this source and constrains its component properties.  
In Section~3 we describe high angular resolution images of {\namesh} obtained with the  
Keck laser guide star adaptive optics (LGS AO) system and NIRC2 camera,
and present limits on the detection of the putative companion.
Section~4 provides additional discussion on the properties of the putative components,
and presents arguments that the system may be relatively old ($\gtrsim$5~Gyr).
Results are summarized in Section~5.

\section{Near-Infrared Spectroscopy}

\subsection{Observations}

Low resolution, near-infrared spectra for {\namesh} were
obtained with IRTF/SpeX on three separate occasions: 
2008 September 8, 2009 June 30, and 2009 November 4 (UT),
as summarized in Table~\ref{tab_obs}.
The SpeX prism mode was used with the 0$\farcs$5 slit 
for all observations, providing 0.7--2.5~$\micron$
coverage in a single order with resolution {\ldl} $\approx 120$
and dispersion of 20--30~{\AA}~pixel$^{-1}$.
To mitigate the effects of differential refraction, the slit was aligned
to the parallactic angle in each observation. 
A0~V stars, either HD~186852  ($V$ = 8.47) or HD~199090 ($V$ = 7.94),  
were observed immediately
after each observation of {\namesh} and at similar airmasses 
for flux calibration and telluric absorption correction.
Internal flat field and argon arc lamp exposures were also obtained 
for pixel response and wavelength calibration.
Data were reduced with the SpeXtool package, version 3.4
\citep{2003PASP..115..389V,2004PASP..116..362C}, using standard settings.  
A detailed description of the reduction
procedures is given in \citet{2007AJ....134.1330B}.

\subsection{The Near-Infrared Spectrum of {\namesh}}

Reduced spectral data of {\namesh} from all three epochs are 
shown in Figure~\ref{fig_nirspec}, 
compared to equivalent data for the L1 spectral standard
2MASS~J14392836+1929149 (hereafter 2MASS~J1439+1929; \citealt{1999ApJ...519..802K}) and 2MASS~J0320$-$0446.
Overall, the near-infrared spectral morphology of {\namesh} is 
similar to these sources and other late-type M and early-type L dwarfs, 
with TiO absorption at red optical wavelengths (0.76 and 0.85~{\micron}); prominent {\wat} absorption
at 1.4 and 1.8~$\micron$; FeH absorption at 0.99,
1.2, and 1.55~$\micron$; unresolved
{\nai} and {\ki} doublet line absorption in the 1.0-1.3~$\micron$ region and at  2.2~$\micron$; and CO bandheads 
at 2.3--2.4~$\micron$.  
{\namesh} has a somewhat redder spectrum than 2MASS~J0320$-$0446 and 2MASS~J1439+1929, consistent with its
redder $J-K_s$ color (1.44$\pm$0.05 versus 1.13$\pm$0.04 and
1.21$\pm$0.03, respectively).  
It also exhibits a somewhat more pronounced 1.27~$\micron$ flux peak.
All three epochs of spectral data of {\namesh} are generally consistent with each other, differing by less than $\pm$5\% across the 0.85--2.4~$\micron$ range.

The one striking feature in the near-infrared spectrum of {\namesh} is the sharp 
dip at 1.6~$\micron$, also seen in the spectrum of 2MASS~J0320$-$0446 but not 
2MASS~J1439+1929.  This feature is nearly coincident with the
1.57--1.64~$\micron$ FeH absorption band but has a different
morphology, forming more of a wedge-shaped divet in contradistinction to the flat plateau
in the spectrum of 2MASS~J1439+1929.  This feature is present in all three
epochs of spectral data.
As discussed in \citet{2007AJ....134.1330B}, this feature can arise in
the combined-light spectrum of a late-type M/L dwarf plus T dwarf binary, where FeH absorption
in the spectrum of the former overlaps with {\meth} absorption in the spectrum of the latter,
resulting in a hybrid feature with this particular shape.  The presence of the 
1.6~$\micron$ dip therefore signifies {\namesh} as a potential unresolved binary system.

\subsection{Spectral Template Analysis}

\subsubsection{Template Sample}

To assess the likelihood of {\namesh} being a binary, we applied the spectral template comparison technique 
described in detail in \citet{2010ApJ...710.1142B}.
A total of 437 SpeX prism spectra of 415 M7--T8 dwarfs were drawn from the SpeX Prism Spectral Libraries\footnote{\url{http://www.browndwarfs.org/spexprism}. Data were drawn specifically from \citet{2004AJ....127.2856B,2006ApJ...639.1095B, 2006ApJ...637.1067B,2007ApJ...658..557B,2008ApJ...681..579B,2008ApJ...674..451B,2004ApJ...604L..61C,2006AJ....131.1007B,2006AJ....131.2722C,2006AJ....132.2074M,2006ApJ...639.1114R,2007ApJ...659..655B,2007AJ....134.1330B,2007ApJ...658..617B,2007ApJ...655..522L,2007AJ....134.1162L,2008ApJ...686..528L,2007ApJ...654..570L,2007AJ....133.2320S,2009AJ....137..304S}; and \citet{2010ApJ...710.1142B}.} and our own observations.  These spectra were required to have median signal-to-noise ratios 
of 20 or greater over the 0.9--2.4~$\micron$ window, and we explicitly excluded known binaries, young cluster objects (e.g., \citealt{2007AJ....134..411M}), and sources specifically
noted to have peculiar spectra associated with low surface gravities, subsolar metallicities, unusual cloud properties or highly uncertain spectral types (e.g.; \citealt{2004AJ....127.3553K, 2006AJ....131.2722C,2007AJ....133..439C,2009AJ....137.3345C,2008ApJ...674..451B,2008ApJ...686..528L}).  

These spectra were interpolated onto a common wavelength scale and flux-calibrated into absolute flux units ($F_{\lambda}$ at 10~pc) using a combination of the 2MASS $M_J$/spectral type relation from \citet{2003AJ....126.2421C} for M7--L2 dwarfs and the 2MASS $M_{K_s}$/spectral type relation from
\citet{2008ApJ...685.1183L} for L2--T8 dwarfs.   Spectrophotometric magnitudes were computed directly from the SpeX spectra by integrating these and a Kurucz model spectrum of Vega with the 2MASS filter profiles (see \citealt{2005ApJ...623.1115C}).  Spectral types were culled from the literature,\footnote{As compiled at \url{http://dwarfarchives.org}.} with optical classifications used for 
M7--L8 dwarfs (tied to the schemes of \citealt{1991ApJS...77..417K} and \citealt{1999ApJ...519..802K}) and near-infrared classifications used for L9--T8 dwarfs (tied to the scheme of \citealt{2006ApJ...637.1067B}) and any M or L dwarf without a published optical classification (based on various schemes: \citealt{2001AJ....121.1710R,2001ApJ...552L.147T,2003ApJ...596..561M,2004ApJ...607..499N,2007ApJ...659..655B}). 

From these flux-calibrated spectra, 99,467 combined-light binary spectral templates were constructed by adding together all possible pairs that satisfied the requirement that one spectrum (the secondary) was of equal or later spectral type than the other (the primary).

\subsubsection{Template Comparison}

Each spectrum of {\namesh} was compared to the single and binary templates to identify best-fit matches.  
All spectra were initially normalized to the 
maximum flux in the 1.2--1.3~$\micron$ region.  We then computed the chi-square difference
statistic between each spectrum of {\namesh} ($S[\lambda]$) and 
each template spectrum ($T[{\lambda}]$):
\begin{equation}
\chi^2 \equiv \sum_{\{ \lambda\} }\left[ \frac{S[{\lambda}]-{\alpha}T[{\lambda}]}{\sigma_c[{\lambda}]} \right]^2.
\label{equ_chisq}
\end{equation}
Here, $\alpha$ is a scaling factor that minimizes {\chisq} (see Equation~2 in \citealt{2008ApJ...678.1372C}),
$\sigma_{c}[{\lambda}]$ is the noise spectrum of {\namesh},
and the sum is performed over the wavelength ranges $\{\lambda\}$ =
0.95--1.35~$\micron$, 1.45--1.8~$\micron$ and 2.0--2.35~$\micron$ in order
to avoid regions of strong telluric absorption.  Unlike \citet{2010ApJ...710.1142B}, we did not use a weighting scheme so each spectral pixel has equal weight in the fit.

Because we are comparing the spectra of distinct sources rather than an optimized spectral model, our expectation is not to achieve {\chisq} $\approx$ 1 for our best-fit cases.  Rather, we wish to assess whether any binary templates provide significantly better fits to the spectra of {\namesh} than the best-fit single template.
As the number of binary templates vastly outnumbers the number of single templates, a better fit (lower {\chisq}) is almost assured so it is necessary to assess the statistical significance of improvement.  As in \citet{2010ApJ...710.1142B}, we used the one-sided F-test for this purpose, using as our distribution statistic the ratio 
\begin{equation}
\eta_{SB} \equiv \frac{{\rm min}(\{\chi^2_{single}\})/\nu_{single}}{{\rm min}(\{\chi^2_{binary}\})/\nu_{binary}} 
\label{equ_eta}
\end{equation}
where $\nu_{single}$ = $\nu_{binary}$ $\equiv$ $\nu$ = 211 is the degrees of freedom in each fit, equal to the number of data points minus one to account for the relative scaling ($\alpha$) between {\namesh} and each template spectrum. To rule out the null hypothesis---that {\namesh} is not a binary---at the 99\% confidence level (CL), we required $\eta_{SB} > 1.34$.  

Finally, since multiple single and binary templates yield {\chisq} values that are statistically indistinct from each other, we computed mean values and uncertainties for the component parameters (i.e., spectral types and relative brightnesses) using a weighting scheme based on the F-distribution:
\begin{equation}
W_i \propto 1-F(\eta_{i0} \mid \nu,\nu).
\end{equation}
Here, $\eta_{i0}$ $\equiv$ {\chisq}$_i$/min([{\chisq]) is the ratio of chi-square residuals between the best-fit template and the $i^{th}$ template, and $F(\eta_{iB} \mid \nu,\nu)$ is the F-distribution probability distribution function.  Parameter means ($\bar{p}$) and uncertainties ($\sigma_p$) were computed as
\begin{equation}
\bar{p} \equiv \frac{\sum_iW_ip_i}{\sum_iW_i}
\label{equ_meanp}
\end{equation}
and
\begin{equation}
\sigma_p^2 = \frac{\sum_iW_i(p_i-\bar{p})^2}{\sum_iW_i}.
\label{equ_sigmap}
\end{equation}

\subsection{Results}

The best-fitting single and binary spectral templates to the data for {\namesh} are shown Figure~\ref{fig_specfits}, and component parameters are listed in Table~\ref{tab_specfits}.
For each observation, a combined-light spectrum composed of a late-type M/early-type L dwarf primary and mid-type T dwarf secondary provides a statistically significant better match to the spectrum of {\namesh} than the best-fit single template, with confidence levels $>$99.9\% in all three cases.  More importantly, the binary templates
consistently reproduce the 1.6~$\micron$ feature, with the T dwarf secondary contributing roughly 5\% to the pseudo-continuum shortward of this feature.  The addition of a T dwarf secondary also explains the strong 1.27~$\micron$ peak in the spectrum of {\namesh}, contributing up to 10\% of the flux between strong {\wat} and {\meth} bands.
Overall, however, the T dwarf secondary contributes minimally to combined-light spectrum; mean
relative $JHK$ magnitudes on the MKO\footnote{Mauna Kea Observatory filter system; see \citet{2002PASP..114..180T} and \citet{2002PASP..114..169S}.} system range over $\Delta{J}$ = 3.0-3.5, $\Delta{H}$ = 4.0-4.5 and $\Delta{K}$ = 4.5-5.2.

The mean component parameters for the three observations of {\namesh} are in fairly good  agreement with each other,
with the first two spectra indicating an L1 primary and a T5.5-T6.5 secondary, a somewhat
later-type combination than 2MASS~J0320$-$0446 (M8.5+T5; \citealt{2008ApJ...681..579B}).  The 2009 Nov 4 spectrum, however, indicates somewhat earlier-type primary and secondary components, M9.5 + T4.5, although the uncertainty on the latter is much higher.  The estimated relative magnitudes are all in agreement, due in part to their large uncertainties ($\pm$0.2-0.4 for $\Delta{J}$ to $\pm$1.5-1.8 for $\Delta{K}$).  Uncertainty-weighted means and uncertainties for these parameters based on all three observations are listed in Table~\ref{tab_component}.

%In composition, {\namesh} therefore appears to be very similar to the spectral and radial velocity binary 2MASS~J0320$-$0446.

\section{LGS AO Imaging}

\subsection{Observations}

In an attempt to identify the faint companion inferred from the spectral analysis,
{\namesh} was imaged on 2009 August 15 (UT) 
with the sodium LGS AO system \citep{2006PASP..118..297W, 2006PASP..118..310V} and
facility near-infrared camera NIRC2 on the 10m Keck II Telescope.
Conditions were fair with clear skies and below average seeing ($\sim$1\arcsec).  
The narrow field-of-view camera of NIRC2 was utilized, 
providing an image scale of
$9.963\pm0.011$~mas/pixel \citep{2006ApJ...649..389P} over a
$10.2\arcsec \times 10.2\arcsec$ field of view with no rotation.  
Observations were conducted through the 
MKO $J$, $H$, and $K_s$-band filters.  For each filter we employed a simple 3-position dither pattern that avoided the noisy, lower left quadrant of the focal plane array.  Exposure times at each dither position are listed in Table~\ref{tab_obs}.   
The LGS provided the wavefront reference source for AO correction, while 
tip-tilt aberrations and quasi-static changes were measured
contemporaneously by
monitoring the $R=13.3$~mag field star USNO 0602-0932082
\citep{2003AJ....125..984M}, located 44$\arcsec$ away from {\namesh}.
We also obtained observations of the nearby star
2MASS J20282802$-$2844422 ($H$ = 13.01~mag)
in the $H$-band as a point spread function (PSF) calibrator, 
using the $R=12.9$~mag field star USNO 0612-0905762
($\rho$ =  43$\arcsec$) for tip-tilt correction.
Because the PSF of LGS-AO images are known to be highly variable based on the brightness of the tip-tilt reference star, the vector (distance and position angle) to the tip-tilt star, and atmospheric effects (airmass, seeing, etc.), the properties of the tip-tilt star for the PSF calibrator are chosen to be as similar as possible to the properties of the tip-tilt star used for the target.  

Images were reduced using custom scripts.  Sky frames, obtained by the median average of the 3 dithered target exposures, were subtracted from each image.  A dome flat was used to correct for pixel-to-pixel sensitivity differences.  The three images were then registered to the peak of the target's PSF, and median combined to produce the final stacked images shown in Figure~\ref{fig_ao}.  No further astrometric or photometric calibration was performed on the final stacked image.

\subsection{Limits on the Presence of a Faint Companion}

Initial examination of the AO images showed no discernible companion to {\namesh} within 3$\arcsec$ of its peak brightness in any of the three bands.  A second source is detected in the field at a separation of 3.6$\arcsec$ and position angle 254$\degr$ which is $\Delta{H} \approx$ 4.5~mag fainter, comparable to the predicted relative brightness of the expected companion.  However, this source is too widely separated to have fallen in the slit with {\namesh} in the SpeX spectral observations, and as such could not have contributed to the spectrum shown in Figure~\ref{fig_nirspec}.  It is also too wide to have been coincident with {\namesh} at the time of the SpeX observations, based on the latter's measured proper motion (350$\pm$15~mas~yr$^{-1}$; \citealt{2008MNRAS.384.1399J}).
%This source shows no optical counterpart in UKST photographic plate images, but this is not unexpected given its faint near-infrared brightness ($H$ $\approx$ 18.5).
A second epoch observation of the {\namesh} field would determine if the faint detected 
source is physically bound to {\namesh}; however, current evidence indicates that it is an unrelated background source.  No other sources are detected in the field with a $\Delta{H}<7$ mag (the 10-$\sigma$ limit).

A better limit on the presence of a close companion to {\namesh} was made from PSF subtraction.
Figure~\ref{fig_psfsub} compares the PSF of {\namesh} to that of the PSF star 2MASS J20282802$-$2844422 at $H$-band, as well as the resulting PSF-subtracted image.  While the calibrator PSF is not a perfect match to that of {\namesh}, the benefits for reducing the halo of {\namesh} are quite evident.  At 0$\farcs$25 the flux from the primary's halo is reduced by a factor of 10, thus facilitating the detection of companions at the expected brightness of the secondary 
beyond that separation.  However, no companions are seen. 

To quantify our detection limits, we generated a set of 10,000 synthetic $H$-band images of the {\namesh} field with
implanted companion sources.  These ``false'' companions were reproductions of the {\namesh} PSF, scaled to the expected magnitude difference of the putative companion ($\Delta{H}$ = 4.2~mag) and inserted at random positions over radial distances spanning 0-100 pixels (0--1$\arcsec$) and all position angles.   
Each of the 10,000 synthetic companions were placed on both the original source image and the PSF-subtracted image.
%(the same distance and position angle distribution was used for both base images).  
We used the SExtractor program \citep{1996A&AS..117..393B} to automatically search for the planted companions.  A detection was considered ``good'' if there was a point source found within 2 pixels of the known location of the planted companion.  
These detections were also visually examined to assess the reliability of the SExtractor detection algorithm, which generally did a comparable job of finding companions as visual inspection albeit with somewhat less reliability within 0$\farcs$25.  Therefore, we consider the SExtractor results to be a conservative upper limit for the maximum detectable separation of a binary {\namesh}.

Figure~\ref{fig_sens} shows the ratio of detections to non-detections as a function of radial distance, based on bins 5 pixels in width.
As expected, the planted companions were detected at closer distances in the PSF-subtracted image compared to the non-PSF-subtracted image (Figure~\ref{fig_sens}), and our reported limit is based on the former.
This analysis indicates that a source with a relative brightness as predicted from the spectral template matching results should have been detected $\gtrsim$50\% of the time at separations beyond 0$\farcs$25.   This is also where visual inspection begins to find the majority of the planted companions.   We therefore adopt 0$\farcs$25 as the maximum angular separation of the companion at the epoch of observation, which corresponds to a 
projected separation limit of 9~AU at the estimated distance of this source (36$\pm$5~pc; \citealt{2007AJ....133..439C}).
Note that this is not a particularly restrictive limit; it falls above the peak of the VLM binary separation distribution (7.2$^{+1.1}_{-1.7}$~AU for a logarithmic distribution; \citealt{2007ApJ...668..492A}),
and only 25\% of known VLM binaries have projected separations wider than 9~AU.

\section{Discussion}

The predicted properties of the putative {\namesh} binary are summarized in Table~\ref{tab_component}.
The luminosities of the components, based on their inferred spectral types
and the $M_{bol}$/spectral type relation of \citet{2007ApJ...659..655B}, are  
{\lbol} = $-3.64{\pm}0.10$ and $-5.1{\pm}0.2$, corresponding to masses of 0.077~{\msun} and 0.030~{\msun}
(0.084~{\msun} and 0.072~{\msun}) for an age of 1~Gyr (10~Gyr), based on the evolutionary models
of \citet{1997ApJ...491..856B}.
With these parameters, the separation limit of the secondary corresponds to 
an orbital period of 82~yr (68~yr) and a primary radial velocity (RV) variation amplitude 
of 27~{\ms} (42~{\ms}), assuming an edge-on circular orbit contained within 9~AU.  This RV amplitude is well below the $\approx$320~{\ms} systematic uncertainties found by \citet{2008ApJ...678L.125B} for their observations of  2MASS~J0320$-$0446, and the 200--300~{\ms} uncertainty found by  \citet{2009A&A...505L...5Z} and \citet{2010ApJ...711.1087K} in comparable near-infrared observations of other VLM sources.  However, 
%considerable effort is now underway to improve near-infrared radial-velocity precision to the 10~{\ms} level (e.g., \citealt{2009arXiv0912.2643F}), and 
if the {\namesh} system is more closely separated than the AO limits indicate---and older---there is a reasonable chance of measuring RV variability in this system in a viable timescale.  For example, for an edge-on orbit with separation of 1~AU and system age of 5~Gyr, the {\namesh} primary would exhibit a 2.5~yr, 240~{\ms} peak-to-peak RV variation, potentially detectable with current instrumentation.
%A more stringent separation limit would make the case for such an RV variation more compelling.

There are several lines of evidence that suggest that {\namesh} is a relatively old VLM system.
The optical spectrum of the primary from \citet{2007AJ....133..439C} shows no evidence for 
the 6708~{\AA} {\lii} absorption line, 
indicating a minimum mass for this component 
of 0.06~{\msun} \citep{1997A&A...327.1039C, 1998ApJ...497..253U} and a
minimum age of 0.4~Gyr \citep{1997ApJ...491..856B}.
This spectrum also shows no discernible H$\alpha$ emission, a feature
commonly observed in younger, more magnetically active late-type 
M and L dwarfs (e.g., \citealt{2007AJ....133.2258S,2008AJ....135..785W}).
%The proper motion of {\namesh} indicates a tangential velocity ({\vtan} = 60$\pm$9~{\kms}) that is
%relatively large for a field L dwarf (e.g., \citealt{2009AJ....137....1F, 2010arXiv1001.3402S}) 
%and also suggestive of an older age.
The proper motion of {\namesh} indicates a tangential velocity ({\vtan} = 60$\pm$9~{\kms}) that is more than 1 sigma faster than the 20~pc sample of L0-L9 dwarfs from \citep{2009AJ....137....1F}.  Given the inferred age of that sample (3.2$^{+1.1}_{-0.9}$~Gyr), the age of {\namesh} can be estimated to be at least 3~Gyr, but more likely older.
Finally, the binary nature of {\namesh} and properties of its components also suggest an older age.
The majority of VLM binaries are found in near-equal mass systems, $q \equiv$ M$_2$/M$_1$ $\approx$ 1
\citep{2003AJ....126.1526B, 2003ApJ...586..512B, 2007ApJ...668..492A}.
Although many imaging surveys are sensitive to mass ratios below $\sim$0.4 for separations smaller than 0$\farcs$2 (e.g., \citealt{2006ApJS..166..585B}), roughly 80\% of known VLM pairs have $q \gtrsim 0.75$.  
Based on the inferred masses of its components, the {\namesh} system would have to 
have an age $\gtrsim$5~Gyr to follow the same trend.  
Direct mass measurements of the components, perhaps through the detection of RV
variations, would provide a more precise constraint on the
age of this system, as has been demonstated for 2MASS~J0320$-$0446 \citep{2009AJ....137.4621B}.

\section{Conclusions}

We have identified {\namesh} as a L0.5 plus T6 spectral binary, based on the presence 
of a peculiar 1.6~$\micron$ feature in its near-infrared spectrum. We associate this feature with overlapping FeH
and {\meth} absorption from the primary and secondary, respectively.
LGS AO imaging failed to resolve a companion to a minimum separation limit of 0$\farcs$25, 
or 9~AU in projected separation,
a fairly unrestrictive limit given the properties of VLM binaries at large.
The inability to resolve this pair may be due in part to the large inferred 
magnitude difference of its components; its
predicted $\Delta{H}$ = 4.2$\pm$0.6~mag is comparable to the largest relative 
magnitudes measured for any VLM pair 
(e.g., 2MASS~J1207334$-$393254AB, $\Delta{H} \approx 5.6$; \citealt{2004A&A...425L..29C};  
GJ~802AB, $\Delta{H} \approx 4.7$; \citealt{2006ApJ...650L.131L};
SCR~J1845$-$6357, $\Delta{H} \approx 4.2$; \citealt{2006ApJ...641L.141B}).
Alternately, {\namesh}, like 2MASS~J0320$-$0446, may be a very tightly bound system,
raising the distinct possibility that direct mass measurements via astrometric and/or radial velocity
orbital monitoring are viable. 
As few direct mass measurements are currently available
for VLM stars and brown dwarfs, additional follow-up observations---e.g., LGS AO aperture masking observations \citep{2006SPIE.6272E.103T} or high resolution spectroscopic monitoring---to resolve this 
system and/or measure the
properties of its components is warranted.
%including LGS AO aperture masking observations (e.g., \citealt{2006SPIE.6272E.103T, 2008ApJ...681..579B}) and high resolution spectroscopic monitoring to search for RV variations.  
{\namesh} joins a growing list of unresolved, VLM binary candidates whose inferred components
straddle the two lowest-luminosity classes of stars and brown dwarfs.

%Given the apparent failure to resolve the system using AO direct imaging techniques at Keck, one possible follow-up observation would be to attempt Aperture Masking to increase the detectability of companions at close radial distances from the primary.  The limiting factor for using Aperture Masking would not be the magnitude difference between the components, but rather the brightness of {\namesh} itself.   All previous reports of Aperture Masking used Natural Guide Star observations because the target was bright enough to drive the AO system.  As discussed above, observations of {\namesh} require the use of the LGS at Keck.  The effectiveness of aperture masking while using the LGS system and an off-axis tip-tilt reference star have not been robustly tested.  Therefore, it is possible that even with aperture masking, the companion will not be resolved.

%EMPHASIZE IMPORTANCE OF FOLLOW-UP

%Given the very close inferred separation of this binary, it is a prime candidate for follow-up radial velocity observations.  By extensive monitoring, an upper limit to the mass ratio can be found, providing a crucial data point for the theoretical models.

\acknowledgements

The authors acknowledge telescope operators Bill Golisch, Paul Sears and Eric Volquardsen, 
and instrument specialist John Rayner at IRTF; 
and Hien Tran and Jason McIlroy at Keck,
for their assistance during the
observations.  We also wish to thank the referee, Kevin Luhman, for his very prompt and helpful review.  
This publication makes use of data from the Two Micron All Sky Survey, which is a joint project of the University of Massachusetts and the Infrared Processing and Analysis Center, and funded by the National Aeronautics and Space Administration and the National Science Foundation. 2MASS data were obtained from the NASA/IPAC Infrared Science Archive, which is operated by the Jet Propulsion Laboratory, California Institute of Technology, under contract with the National Aeronautics and Space Administration.
This research has benefitted from the M, L, and T dwarf compendium housed at \url{http://DwarfArchives.org} and maintained by Chris Gelino, Davy Kirkpatrick, and Adam Burgasser;
the Very-Low-Mass Binaries Archive housed at \url{http://www.vlmbinaries.org} and maintained by Nick Siegler, Chris Gelino, and Adam Burgasser; and
the SpeX Prism Spectral Libraries, maintained by Adam Burgasser at 
\url{http://www.browndwarfs.org/spexprism}.
The authors recognize and acknowledge the 
very significant cultural role and reverence that 
the summit of Mauna Kea has always had within the 
indigenous Hawaiian community.  We are most fortunate 
to have the opportunity to conduct observations from this mountain.

Facilities: \facility{IRTF~(SpeX)}, \facility{Keck~(NIRC2,LGS)}

\clearpage

\begin{figure}
\epsscale{0.8}
\plotone{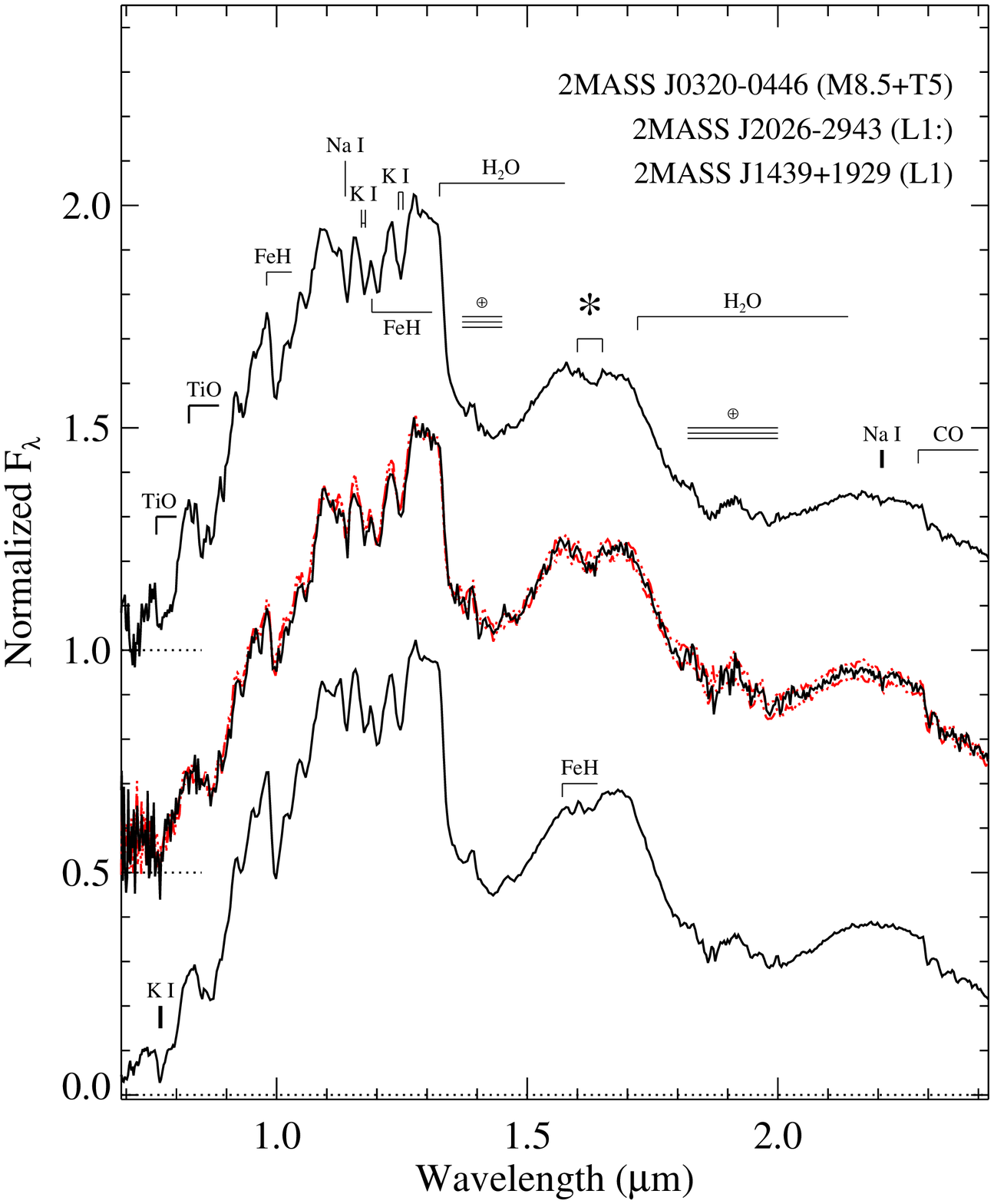}
\caption{SpeX prism spectrum of {\namesh} (center: black line from 2009 June 30 UT; red lines from 2008 September 8 UT and 2009 November 4 UT) compared
to equivalent data for the M8.5+T5 spectral binary
2MASS~J0320$-$0446 (top; data from \citealt{2008ApJ...681..579B})
and the L1 spectral standard 
2MASS~J1439+1929 (bottom; data from \citealt{2004AJ....127.2856B}).
All three spectra
are normalized at 1.25~$\micron$ and offset by
constants (dotted lines).  Prominent features resolved by these
spectra are indicated. The peculiar 1.6~$\micron$ feature in the spectra
of 2MASS~J0320-0446 and {\namesh} as discussed in the text 
is indicated by an asterisk.
\label{fig_nirspec}}
\end{figure}

\clearpage

\begin{figure}
\centering
\includegraphics[width=0.42\textwidth]{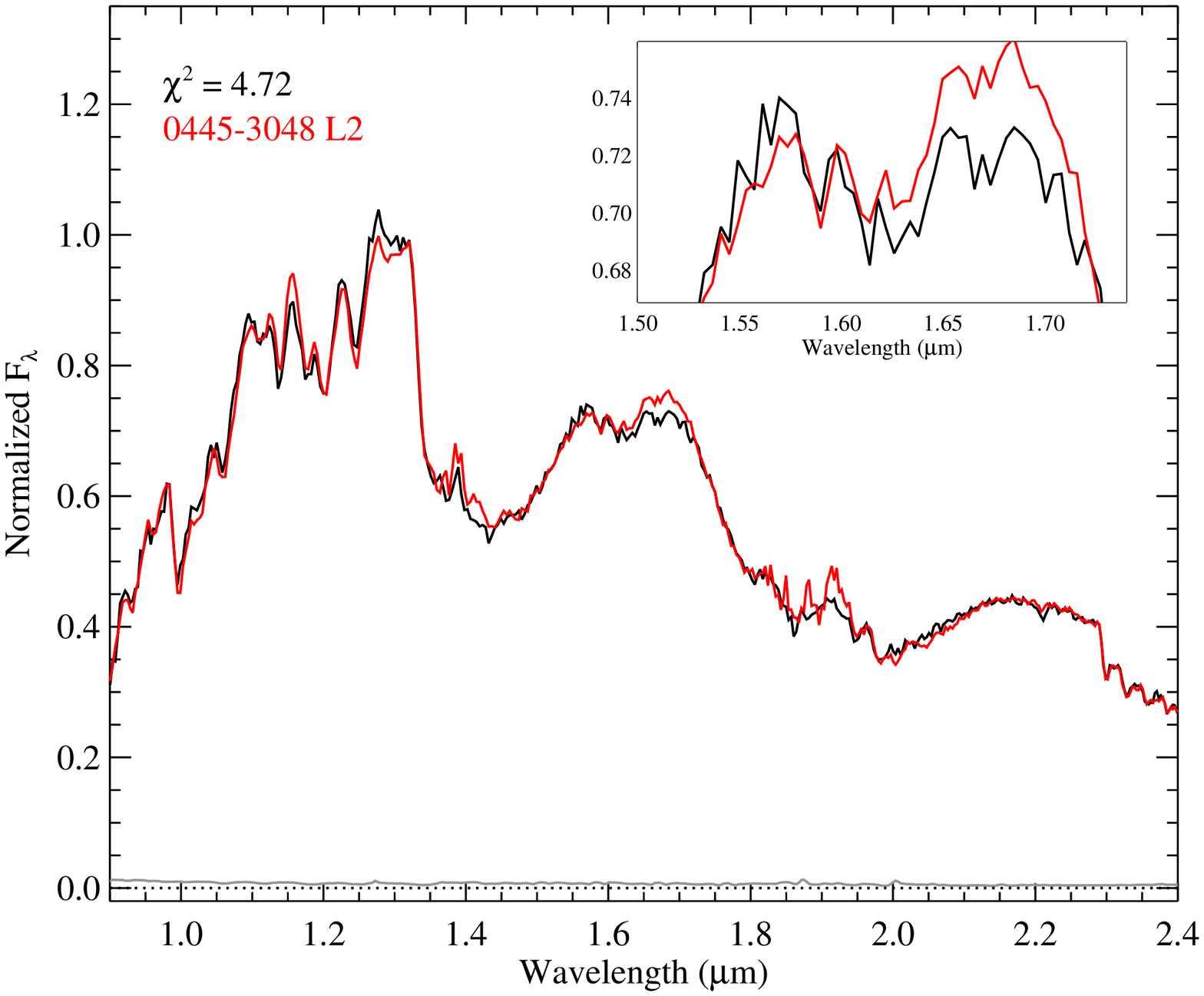}
\includegraphics[width=0.42\textwidth]{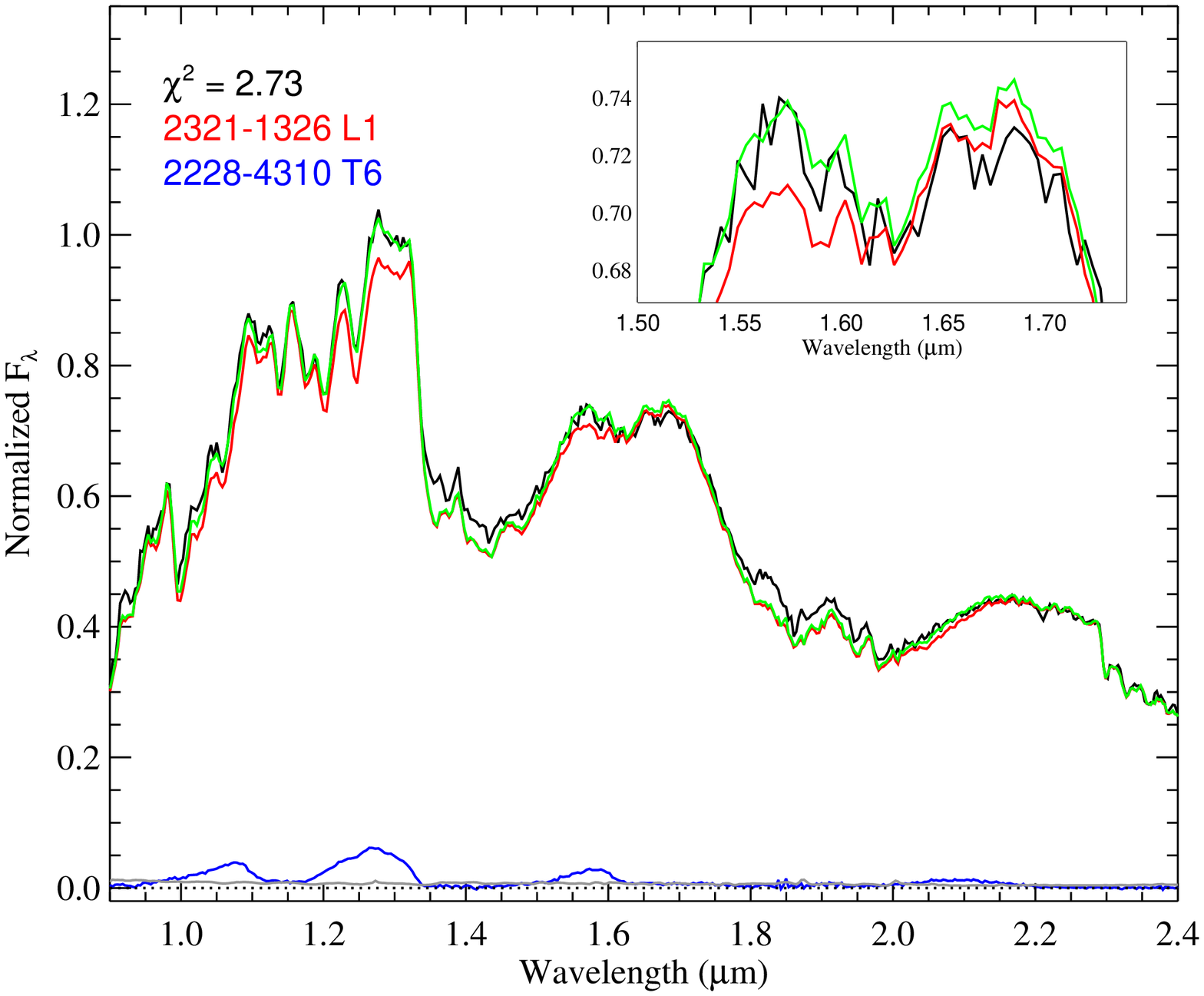} \\
\includegraphics[width=0.42\textwidth]{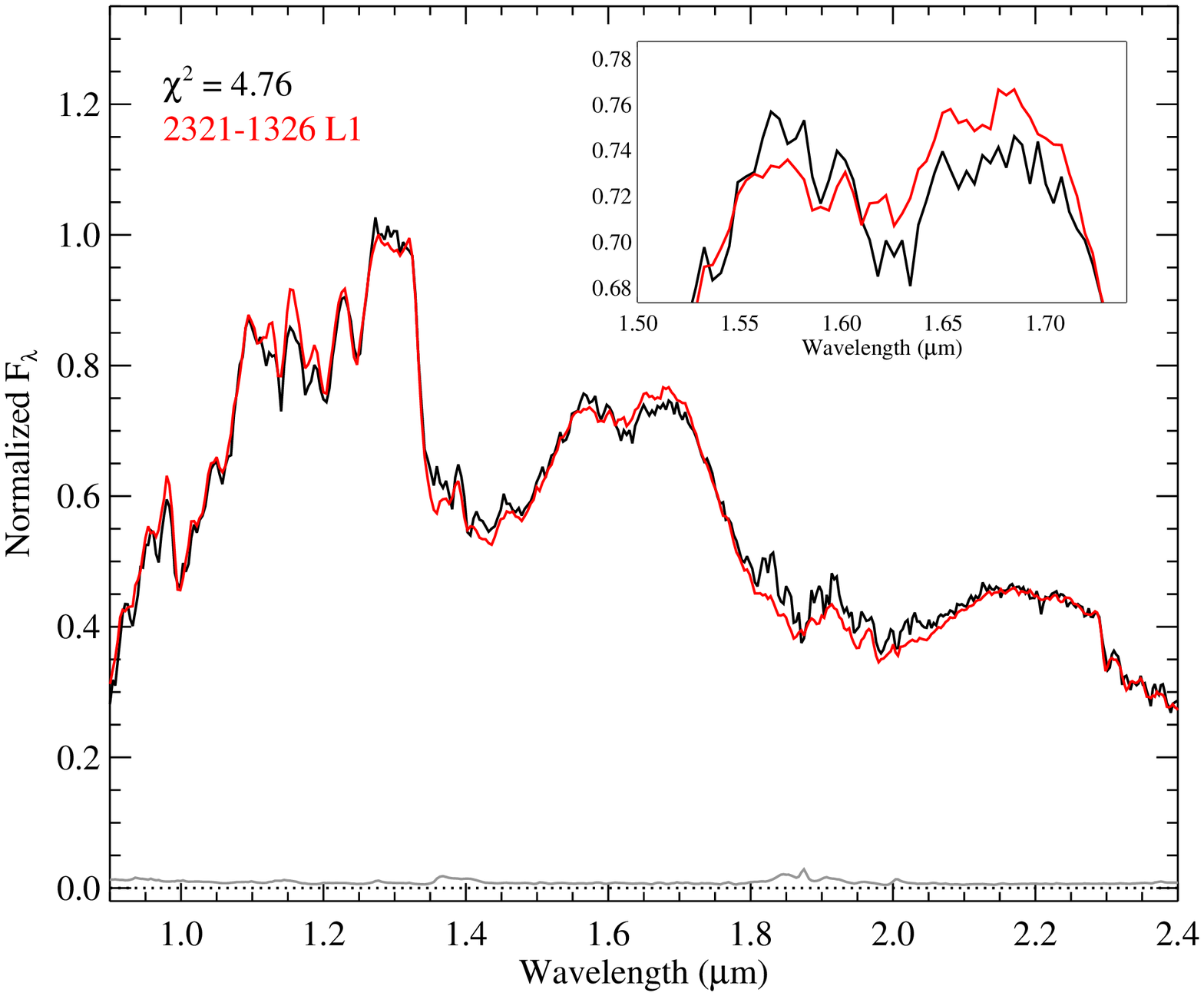}
\includegraphics[width=0.42\textwidth]{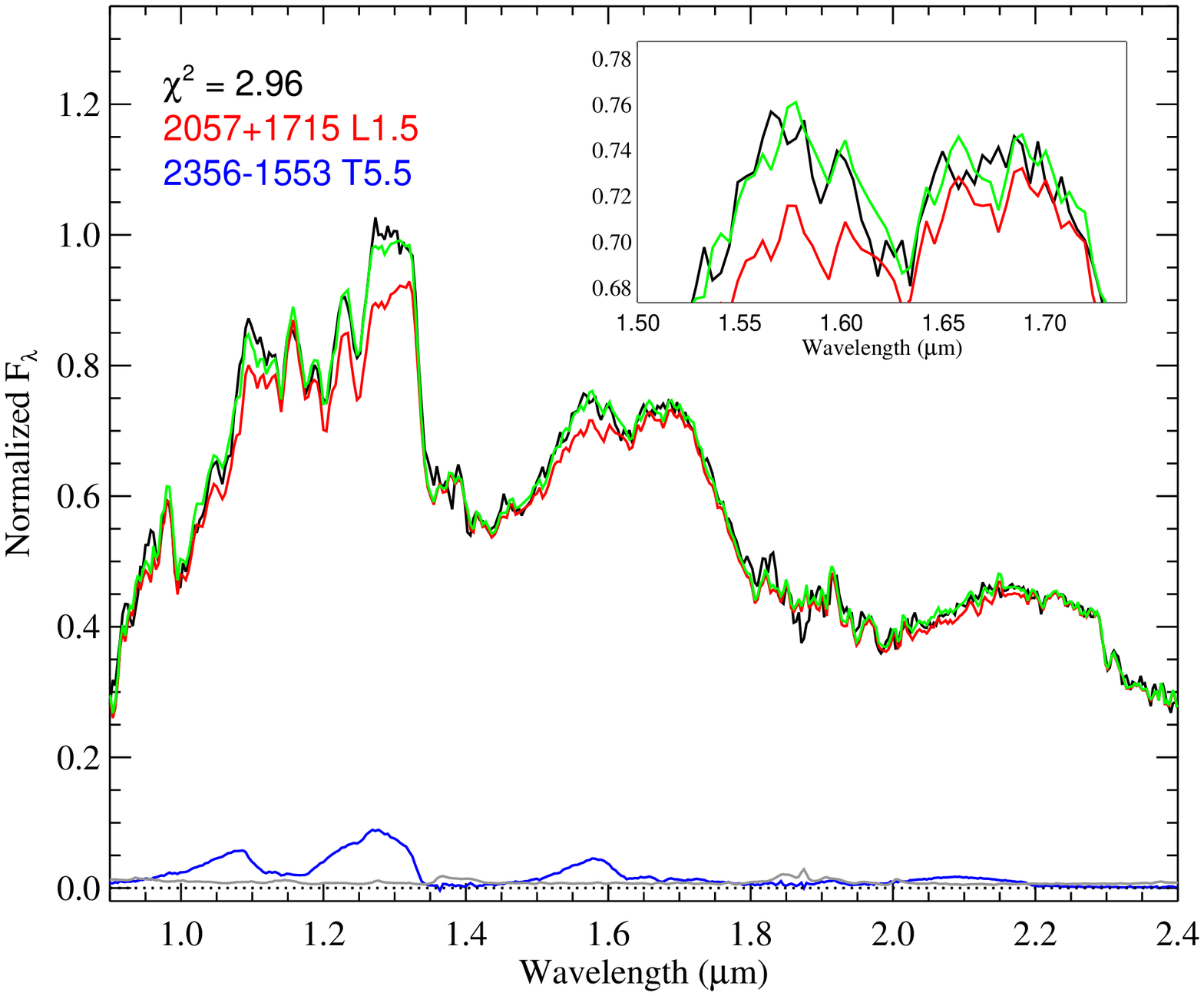} \\
\includegraphics[width=0.42\textwidth]{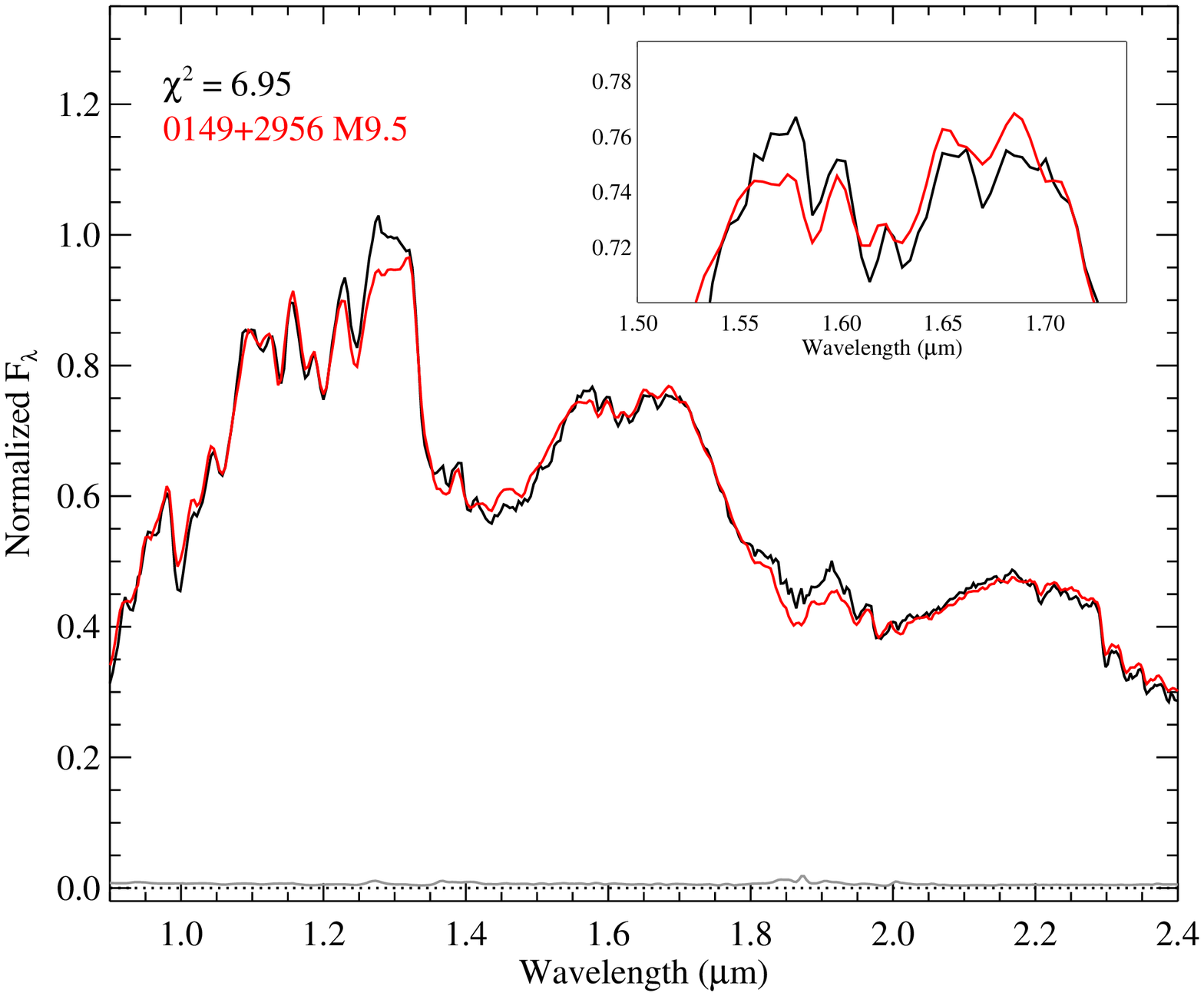}
\includegraphics[width=0.42\textwidth]{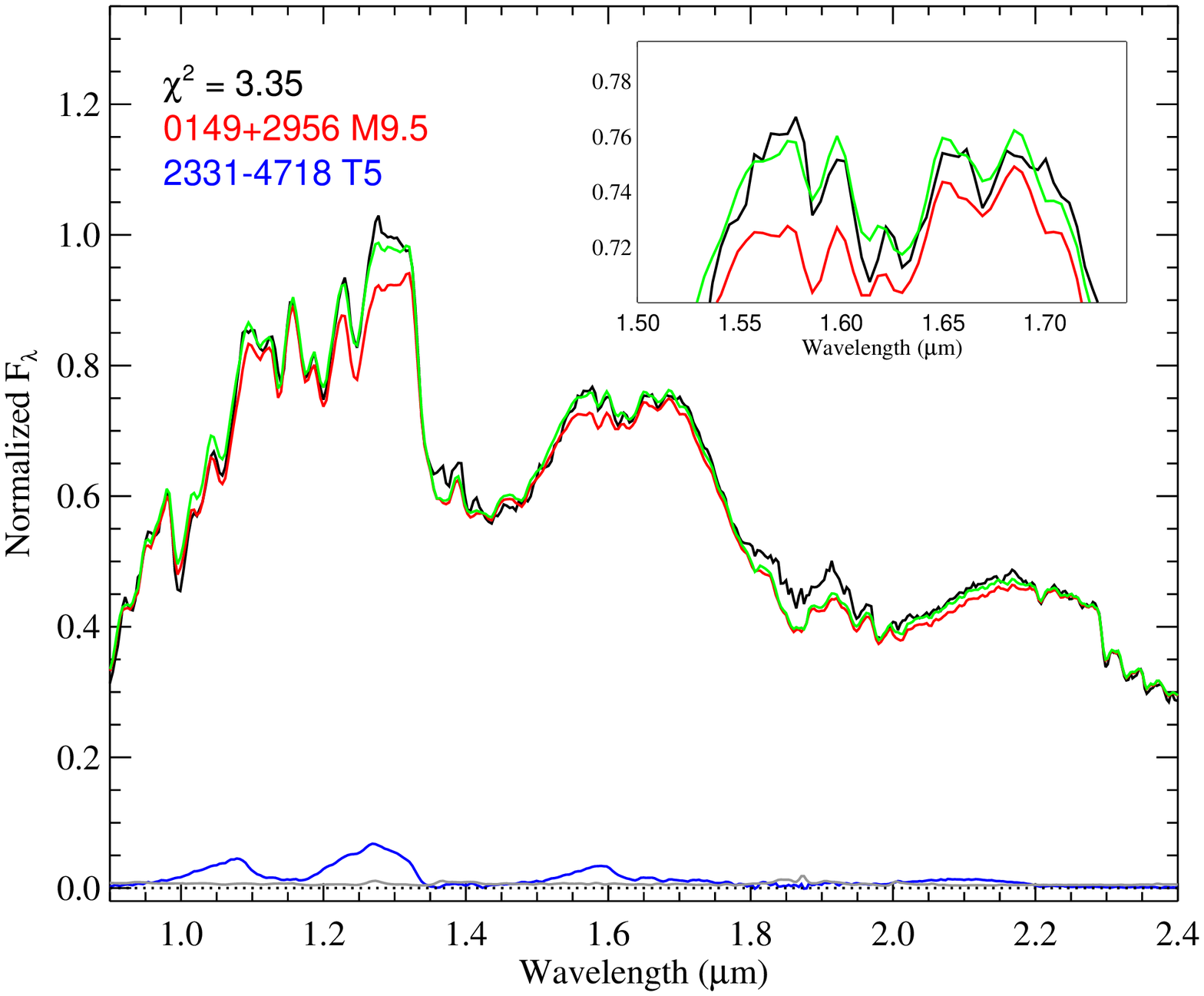} \\
\caption{(Left): The best-fit single spectral templates (red lines) to the three epochs of spectral data for {\namesh} (black lines), from top to bottom: 2008 September 8, 2009 June 30 and 2009 November 4 (UT). (Right): The best-fit binary spectral templates (green lines), and component primaries (red lines) and secondaries (blue lines) for the same epochs.
All spectra are normalized in the 1.2--1.3~$\micron$
region, with the single and binary templates scaled to minimize
their {\chisq} deviations.  The primary and secondary component spectra
in the bottom panel 
are scaled according to their contribution to the
binary templates.  
Inset boxes show a close-up of the 1.5--1.75~$\micron$ region
where the peculiar 1.6~$\micron$ feature is located. 
\label{fig_specfits}}
\end{figure}

\clearpage

\begin{figure}\centering
\includegraphics[width=0.32\textwidth]{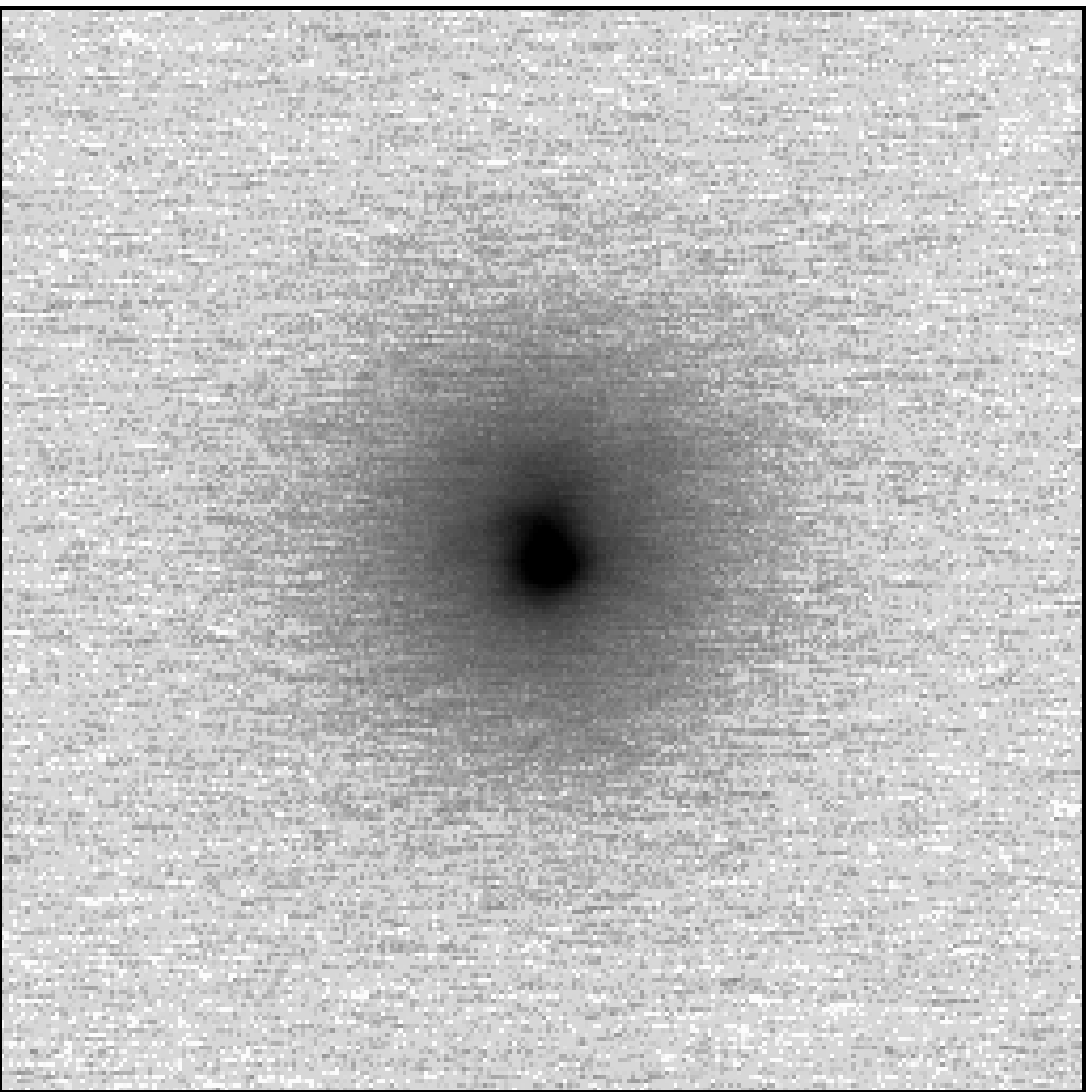}
\includegraphics[width=0.32\textwidth]{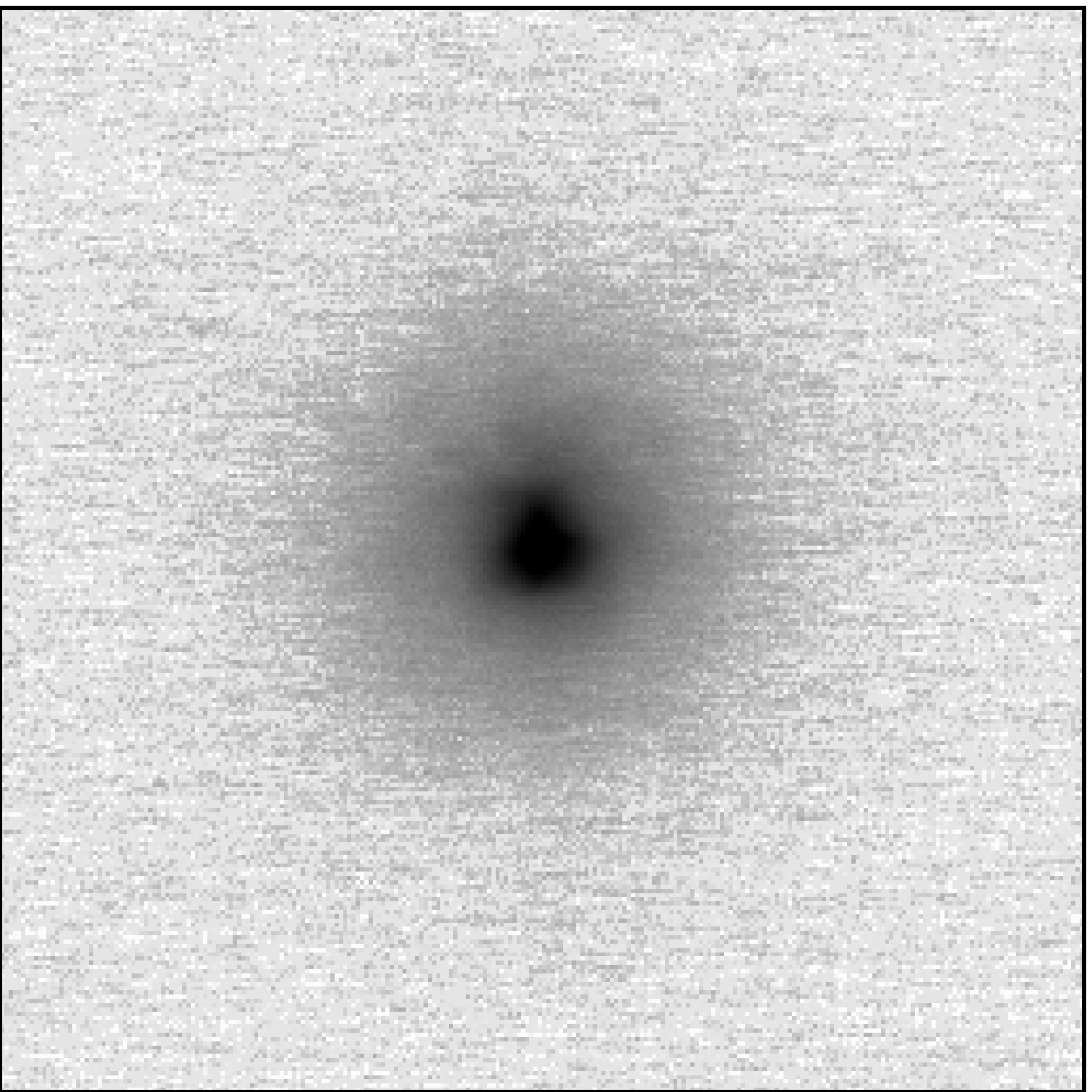}
\includegraphics[width=0.32\textwidth]{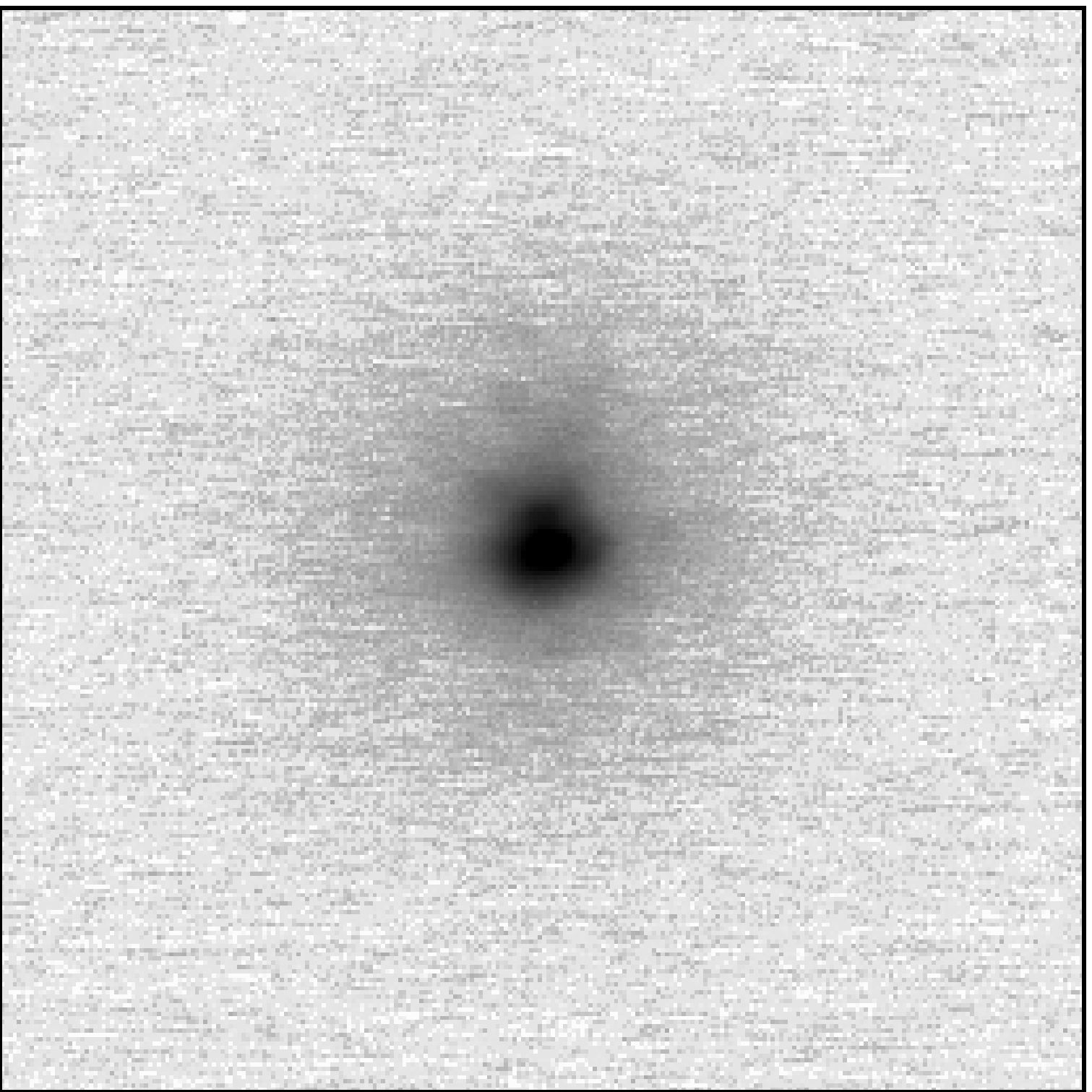}
\caption{$J$,$H$,$K_s$ Keck NIRC2/LGS AO images of {\namesh} from 2009 August 15 (UT).  
The images are 2$\farcs$5 on a side and oriented with North up and East to the left.  The full-width 
half maxima and Strehl ratios for the final images are 81mas and 1.4\% for $J$, 83mas and 4.9\% 
for $H$, and 84mas and 12\% for $K_s$. 
\label{fig_ao}}
\end{figure}

\clearpage

%\begin{figure}
%\epsscale{1.0}
%\plottwo{im_invert.eps}{psf_invert.eps}
%\caption{PSF-subtracted image of {\namesh} at $H$.  The images are 2.5\arcsec\ on a side and have the same display parameters (min/max values, stretch, etc.). 
%\label{fig_psfsub}}
%\end{figure}

%\clearpage

\begin{figure}
%\epsscale{0.8}
\plotone{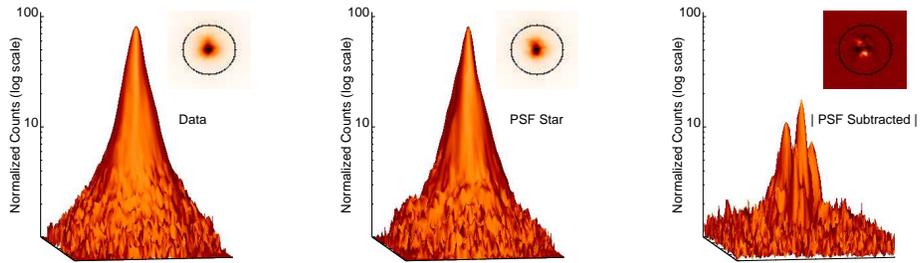}
\caption{PSF-fitting to $H$-band imaging data for {\namesh}. Shown are the normalized surface fluxes on a logarithmic scale and images on a linear scale (inset boxes) for 0$\farcs$8$\times$0$\farcs$8 regions around the source (left), the PSF calibrator star (middle) and the PSF-subtracted residuals (absolute values) as scaled to the original data (right).  Maximum residuals are roughly 20\% of the peak source flux.  The 0$\farcs$25 radius region around the source position beyond which a companion can be ruled out is indicated in the inset images.   At the displayed scaling, the inferred secondary of this system would have a peak flux of roughly 2.  
\label{fig_psfsub}}
\end{figure}

\clearpage

\begin{figure}
\epsscale{0.9}
\plotone{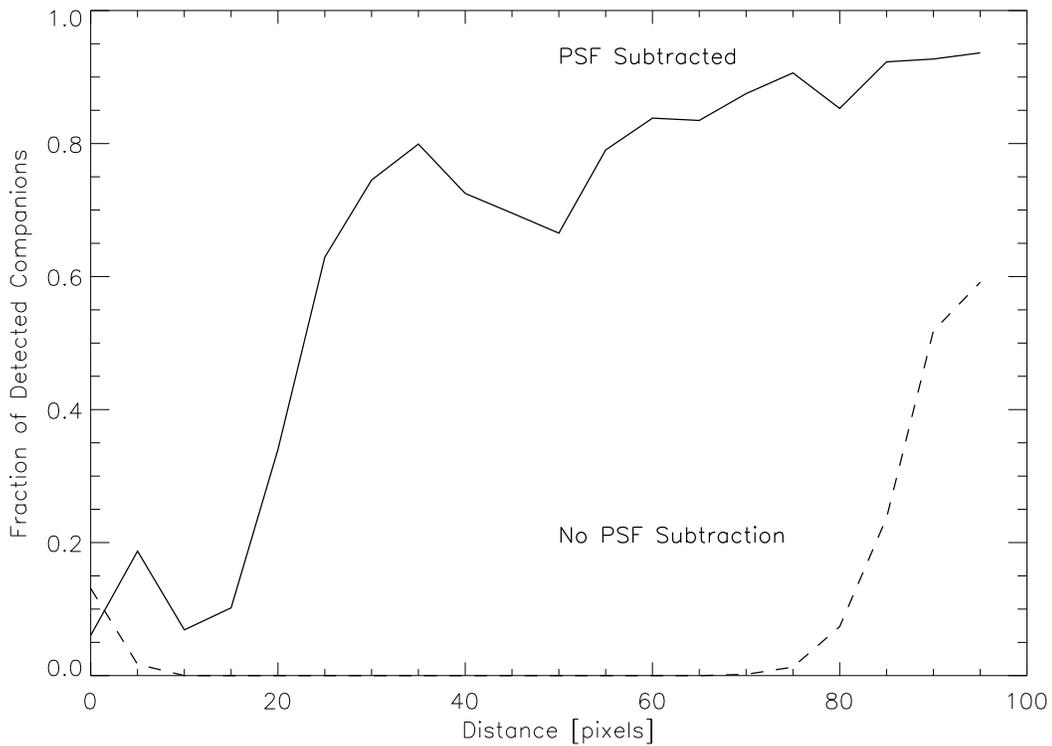}
\caption{The probability of detecting a companion to {\namesh} comparable in brightness to that inferred from the spectral analysis in the PSF-subtracted (solid line) and non-PSF-subtracted (dashed line) images. The benefits of using PSF subtraction for companion detection at small radial distances is clearly evident in the greatly enhanced fraction of detected companions closer than 80 pixels ($\approx$0$\farcs$8).  We adopt a minimum companion detection limit of 0$\farcs$25, where the probability of detection rises above 50\%. \label{fig_sens}}
\end{figure}

\clearpage

\begin{deluxetable}{llllll}
\tabletypesize{\footnotesize}
\tablecaption{Observations of {\namesh}. \label{tab_obs}}
\tablewidth{0pt}
\tablehead{
\colhead{Instrument} &
\colhead{Date} &
\colhead{Setup} &
\colhead{Integration} &
\colhead{Airmass}  &
\colhead{Conditions}  \\
}
\startdata
IRTF/SpeX  & 2008 Sep 8 & 0$\farcs$5 slit, prism mode & 150s$\times$6 & 1.54 & clear, 1$\arcsec$ seeing \\
IRTF/SpeX & 2009 Jun 30  & 0$\farcs$5 slit, prism mode & 120s$\times$8 & 1.54 & clear, 1$\arcsec$ seeing \\
IRTF/SpeX & 2009 Nov 4 & 0$\farcs$5 slit, prism mode & 150s$\times$8 & 1.60 & light cirrus, 0$\farcs$5 seeing \\
Keck/NIRC2 + LGSAO & 2009 Aug 15 & $J$, narrow camera & 60s$\times$4 & 1.58 & clear, 1$\arcsec$ seeing, RH$\sim$20\% \\
Keck/NIRC2 + LGSAO & 2009 Aug 15 & $H$, narrow camera & 30s$\times$6 & 1.54 & \nodata \\
Keck/NIRC2 + LGSAO & 2009 Aug 15 & $K_s$, narrow camera & 15s$\times$16 & 1.55 & \nodata \\
\enddata
\end{deluxetable}

\begin{deluxetable}{llllcccl}
\tabletypesize{\scriptsize}
\tablecaption{Spectral Template Fits. \label{tab_specfits}}
\tablewidth{0pt}
\tablehead{
\colhead{Primary} &
\colhead{SpT} &
\colhead{Secondary} &
\colhead{SpT} &
\colhead{$\Delta{J}$}  &
\colhead{$\Delta{H}$}  &
\colhead{$\Delta{K}$}  &
\colhead{\chisq/CL}  \\
}
\startdata
\cline{1-8}
\multicolumn{8}{c}{2008 Sep 8} \\
\cline{1-8}
2MASSI J0445538-304820 & L2 & \nodata & \nodata &  \nodata & \nodata & \nodata & 4.72 \\
2MASS J23211254-1326282 & L1 & 2MASS J22282889-4310262 & T6 & 3.37 & 4.39 & 4.85 & 2.73 \\ 
$<$Primary$>$ & L1.2$\pm$0.5 & $<$Secondary$>$ & T6.5$\pm$0.7 & 3.5$\pm$0.4 & 4.5$\pm$1.1 & 5.2$\pm$1.8 & 99.996\% \\
\cline{1-8}
\multicolumn{8}{c}{2009 Jun 30} \\
\cline{1-8}
2MASS J23211254-1326282 & L1 & \nodata & \nodata &  \nodata & \nodata & \nodata & 4.76 \\
2MASSI J2057153+171515 & L1.5 & 2MASSI J2356547-155310 & T5.5 & 2.85 & 3.83 & 4.39 & 2.96 \\ 
$<$Primary$>$ & L0.8$\pm$0.9 & $<$Secondary$>$ & T5.5$\pm$0.6 & 3.0$\pm$0.2 & 4.0$\pm$1.1 & 4.6$\pm$1.7 & 99.971\% \\
\cline{1-8}
\multicolumn{8}{c}{2009 Nov 4} \\
\cline{1-8}
2MASS J01490895+2956131 & M9.5 & \nodata & \nodata &  \nodata & \nodata & \nodata & 6.95 \\
2MASS J01490895+2956131 & M9.5 & 2MASS J23312378-4718274 & T5 & 3.23 & 4.08 & 4.60 & 3.35 \\ 
$<$Primary$>$ & M9.5$\pm$0.0 & $<$Secondary$>$ & T4.7$\pm$1.2 & 3.2$\pm$0.3 & 4.0$\pm$1.0 & 4.5$\pm$1.5 & 99.999\% \\
\enddata
\end{deluxetable}

\begin{deluxetable}{lcccl}
\tabletypesize{\footnotesize}
\tablecaption{Component and System Parameters for {\namesh}. \label{tab_component}}
\tablewidth{0pt}
\tablehead{
\colhead{Parameter} &
\colhead{{\namesh}A} &
\colhead{{\namesh}B} &
\colhead{Difference} &
\colhead{Ref} \\
}
\startdata
Spectral Type & L0.5 & T6.0 & \nodata & 1 \\
${J}$\tablenotemark{a} (mag) & 14.86$\pm$0.03  &  17.98$\pm$0.15 & 3.12$\pm$0.16 & 1 \\
${H}$\tablenotemark{a} (mag) & 13.97$\pm$0.03  &  18.2$\pm$0.6 & 4.2$\pm$0.6 & 1 \\
${K}$\tablenotemark{a} (mag) & 13.37$\pm$0.04  &  18.1$\pm$1.0 & 4.7$\pm$1.0 & 1 \\
{\lbol}\tablenotemark{b} & -3.64$\pm$0.10 & -5.1$\pm$0.2 & 1.5$\pm$0.3 & 1,2 \\
$d$ (pc) & 36$\pm$5 & \nodata & \nodata & 3 \\
$\mu$ (mas/yr) & 351$\pm$15 & \nodata & \nodata & 4 \\
{\vtan} ({\kms}) & 66$\pm$7 & \nodata & \nodata & 3,4 \\
$\rho$ (AU) & $<$0$\farcs$25 ($<$9~AU) & \nodata & \nodata & 1,3 \\
%Mass\tablenotemark{e} (M$_{\sun}$) at 0.5 Gyr & 0.065 & 0.021 &  0.33\tablenotemark{f} \\ 
Mass\tablenotemark{c} (M$_{\sun}$) at 1 Gyr & 0.077 & 0.030 &  0.39\tablenotemark{d} & 5 \\
Mass\tablenotemark{c} (M$_{\sun}$) at 5 Gyr & 0.084 & 0.063 &  0.75\tablenotemark{d} & 5 \\
Mass\tablenotemark{c} (M$_{\sun}$) at 10 Gyr & 0.084 & 0.072 &  0.86\tablenotemark{d} & 5 \\
\enddata
\tablenotetext{a}{Synthetic magnitudes on the MKO system, based on 2MASS $JHK_s$ photometry for the unresolved source and spectrophotometry of the binary template components.}
\tablenotetext{b}{Based on the $M_{bol}$/spectral type relation of \citet{2007ApJ...659..655B}.}
\tablenotetext{c}{Based on evolutionary models from \citet{1997ApJ...491..856B} and estimated luminosities.}
\tablenotetext{d}{Mass ratio $q \equiv$ M$_2$/M$_1$.}
\tablerefs{(1) This paper; (2) \citet{2007ApJ...659..655B}; (3) \citet{2007AJ....133..439C}; (4) \citet{2008MNRAS.384.1399J}; (5) \citet{1997ApJ...491..856B}.}
\end{deluxetable}

\clearpage

%\bibliography{../biblibrary}

\end{document}